\documentclass[final,5p,times,twocolumn]{elsarticle}

\usepackage{amssymb}
\usepackage{amsmath}
\usepackage{hyperref}
\usepackage{xr-hyper} % Must be loaded after hyperref
\usepackage{graphicx}
\usepackage{multirow}
\usepackage{soul}
\usepackage{subcaption} 
\usepackage{cuted}
\usepackage[none]{hyphenat}
\usepackage{subfiles}
\usepackage{verbatim}
\usepackage{romannum}
\usepackage{caption}

\journal{Surfaces and Interfaces}

\begin{document}

\begin{frontmatter}



\title{Atomic-Scale Mechanisms of SiO$_2$ Plasma-Enhanced Chemical Vapor Deposition Revealed by Molecular Dynamics with a Machine-Learning Interatomic Potential} %% Article title

%% author information
\author[SNU]{Jaehoon Kim}
\author[SNU]{Minseok Moon}
\author[SNU]{Hyunsung Cho}
\author[SD]{Hyeon-Deuk Kim}
\author[SD]{Rokyeon Kim}
\author[SD]{Gyehyun Park}
\author[SNU,KIAS]{Seungwu Han}
\author[INU]{Youngho Kang\corref{cor1}}
\ead{youngho84@inu.ac.kr}

%% corresponding author footer text
\cortext[cor1]{Corresponding author}

%% affiliation information
\affiliation[SNU]{organization={Department of Materials Science and Engineering and Research Institute of dvanced Materials, Seoul National University},
  city={Seoul},
  postcode={08826},
  country={Republic of Korea}}
  
\affiliation[SD]{organization={Display Research Center, Samsung Display Company Ltd.},
  city={Yongin-si},
  postcode={17113},
  country={Republic of Korea}}

\affiliation[KIAS]{organization={Korea Institute for Advanced Study},
  city={Seoul},
  postcode={02455},
  country={Republic of Korea}}
  
\affiliation[INU]{organization={Department of Materials Science and Engineering, Incheon National University},
  city={Incheon},
  postcode={22012},
  country={Republic of Korea}}

%% Abstract
\begin{abstract}
Plasma-enhanced chemical vapor deposition (PECVD) of silicon dioxide (SiO$_2$) is widely used for low-temperature fabrication of dielectric thin films, yet its atomic-scale growth mechanisms remain incompletely understood. In this work, we investigate SiO$_2$ PECVD using silane and N$_2$O as source gases via molecular dynamics simulations driven by a machine-learning interatomic potential. By systematically varying the oxidant-to-silane-derived species ratio $r$, we elucidate the evolution of film stoichiometry, density, and hydrogen content. Formation of the Si--O--Si network primarily proceeds via oxidation of surface Si--H groups to form Si--OH species, followed by condensation of neighboring Si--OH groups that produces H$_2$O as the dominant byproduct. At low $r$, H$_2$ formation via reactions between Si--H and Si--OH groups also contributes to the network formation. Increasing oxidant supply promotes the network formation through oxidation of residual Si--H species, suppressing hydrogen incorporation and leading to saturation of the Si/O ratio. Rapid chemisorption of silane-derived species, together with steric hindrance from pre-deposited species, results in localized growth and surface roughness. We further show that high-kinetic-energy plasma species can etch SiO$_2$ films, which potentially limits growth rates and enhances surface roughness under high RF-power conditions. These results provide atomic-scale insight into PECVD growth and guidance for optimizing film composition and quality.
\end{abstract}

%% Keywords

\end{frontmatter}


\section{Introduction}

Silicon dioxide (SiO$_2$) thin films have long played a central role in semiconductor technology, serving as gate dielectrics, interlayer dielectrics, and passivation layers in metal–oxide–semiconductor field-effect transistors (MOSFETs) and thin-film transistors (TFTs) owing to their excellent thermal stability, chemical robustness, and insulating properties \cite{fortunato2012oxide}. Among the various techniques for growing SiO$_2$ thin films, plasma-enhanced chemical vapor deposition (PECVD) offers several advantages, including high deposition rates and compatibility with a wide range of substrate materials, making it highly attractive for cost-effective and large-scale manufacturing \cite{seshan2012handbook,batey1986low}. In particular, PECVD enables SiO$_2$ deposition at relatively low temperatures (typically below 400~$^{\circ}\mathrm{C}$), in contrast to thermal oxidation of silicon, which requires processing temperatures exceeding 800~$^{\circ}\mathrm{C}$ \cite{deal1965general}. This low-temperature capability makes PECVD suitable for substrates that cannot withstand high thermal budgets, such as glass and polymer substrates used in display technologies \cite{fortunato2012oxide}. Low-temperature PECVD is also essential for back-end-of-line (BEOL) device fabrication, where excessive thermal exposure damages previously fabricated underlying devices \cite{maex2003low}.

Despite the various advantages of PECVD for SiO$_2$ thin films, careful control of the deposition process is required to obtain high-quality thin films, because the film properties, such as the Si/O atomic ratio and mass density, strongly depend on deposition conditions including precursor flow ratios, radio-frequency (RF) power, and deposition temperature \cite{adams1981characterization,deenapanray1998characterization,calta2024depth}. In addition, the PECVD process inherently introduces hydrogen into the growing film, often resulting in substantial hydrogen concentrations in SiO$_2$ \cite{zhang2003thermo,martinu2010plasma}. These hydrogen-related species can act as charge traps at the interface with the semiconductor channel in transistors. More critically, hydrogen may diffuse into the active channel during device fabrication and operation, thereby degrading the electrical performance and long-term reliability of oxide-based TFTs \cite{kamiya2010present,kang2015hydrogen,park2025suppressing}. Therefore, a detailed understanding of the atomic-scale growth mechanisms is essential for optimizing PECVD SiO$_2$ film quality and device performance. However, these mechanisms remain incompletely understood so far.

Several studies have previously investigated SiO$_2$ PECVD processes to gain insight into the underlying growth mechanisms. For example, Kushner et al.~\cite{kushner1993plasma} simulated electron kinetics and plasma chemistry in PECVD processes using silane ($\mathrm{SiH_4}$) and nitrous oxide ($\mathrm{N_2O}$), which are widely employed precursor combinations for SiO$_2$ deposition. Their simulations suggested that oxidized silane derivatives, particularly $\mathrm{SiH}_n\mathrm{O}$ species, may play dominant roles in film growth. However, the work primarily focused on gas-phase plasma reactions and did not provide detailed insights into the surface growth mechanisms. Smith et al.~\cite{smith1993chemistry} suggested possible reaction pathways involving a broader range of plasma species, including $\mathrm{SiH}_n$, $\mathrm{SiH}_n(\mathrm{OH})_p$, $\mathrm{SiO}_n$, and atomic oxygen. Nonetheless, they were largely schematic and were not rigorously validated. More importantly, these previous studies offered limited insight into the atomic-scale surface reactions that govern film growth after plasma-species incidence. Recently, a refined growth model based on silanone ($\mathrm{SiH_2O}$) and atomic oxygen as dominant plasma species leading to film growth was proposed~\cite{zhang2024formation}. While this model suggested a reaction pathway leading to the formation of Si--O--Si networks, it remains simplified and does not fully explain key film-quality metrics such as stoichiometry, density evolution, and hydrogen incorporation.

Computational approaches have become effective tools for elucidating material deposition processes at the atomic scale~\cite{sibanda2022review,dang2025decoding}. Among them, density functional theory (DFT) has been widely employed to investigate chemical reactions associated with deposition because of its high accuracy and capability to describe electron transfer and bond rearrangements~\cite{deng2014interaction,fang2015stepwise}. However, the high computational cost of DFT typically restricts its application to static evaluations of individual reactions along predefined reaction pathways. To capture the dynamic evolution of surfaces during deposition without imposing prior assumptions about reaction mechanisms, simulations at spatial and temporal scales beyond the practical capability of DFT are required. Although classical molecular dynamics (MD) can simulate such large systems and long time scales, it generally relies on empirical force fields that are parameterized for specific systems and often assume fixed bonding topologies~\cite{uene2022reactive,shendokar2023evaluating}. Consequently, classical MD is often not well suited to accurately describe the diverse and complex chemical reactions occurring during thin-film deposition processes.

Recently, MD driven by a machine-learning interatomic potential (MLIP) has emerged as a powerful approach that bridges the gap between the accuracy of DFT and the efficiency of classical MD. Trained on high-fidelity quantum mechanical datasets, MLIPs combine near-DFT accuracy with the computational efficiency of classical potentials, enabling large-scale simulations of surfaces, interfaces, and complex chemical processes \cite{hong2023applications,wan2024construction}. In particular, pretrained general-purpose MLIPs based on graph neural networks, such as SevenNet-0~\cite{SevenNet0_2024} and MACE-MP-0~\cite{batatia2025foundation}, have demonstrated strong transferability across diverse chemical environments by learning chemically meaningful representations from extensive datasets spanning broad chemical spaces.

In this work, we investigate SiO$_2$ PECVD using MD simulations driven by an MLIP, focusing on SiH$_4$ and N$_2$O as source gases. We first develop an MLIP tailored for SiO$_2$ PECVD simulations by fine-tuning the pretrained SevenNet-0 (7net-0) model \cite{park2024scalable,SevenNet0_2024}. After thorough validation, the fine-tuned MLIP is employed to perform MD simulations of SiO$_2$ film growth. Our simulations show that increasing the injection ratio of oxidant to silane-derived species in the plasma phase (represented by atomic oxygen and SiH$_2$O, respectively) drives the deposited film toward the ideal stoichiometry of SiO$_2$. The dominant reaction pathway for forming Si--O--Si networks involves oxidation of Si--H species originating from adsorbed SiH$_2$O, followed by condensation of neighboring Si--OH groups that produce H$_2$O as the primary byproduct. Additional reaction pathways that contribute to network formation are also identified. Furthermore, the simulations reveal that insufficient oxidation and condensation lead to significant hydrogen incorporation, which suppresses Si--O--Si network formation and limits film densification. Finally, we show that impacts of high-kinetic-energy reactive species can induce etching of SiO$_2$ films, potentially limiting the net growth rate and increasing surface roughness under high RF-power conditions. 

\section{Methods}
\subsection{DFT calculations}

Spin-polarized DFT calculations are performed using the Vienna Ab initio Simulation Package (VASP) \cite{kresse1996efficiency,kresse1996efficient,kresse1999ultrasoft,kresse1995ab}. The exchange-correlation functional is described by the generalized gradient approximation in the Perdew-Burke-Ernzerhof (PBE)~\cite{perdew1996generalized} method. We employ a plane-wave cutoff energy of 520 eV. We sample a single $\Gamma$-point for Brillouin-zone integration. This setup ensures the convergence of the total energy and atomic force within 1 meV/atom and 0.02 eV/\AA{}, respectively. First-principles molecular dynamics (FPMD) simulations are performed in the NVT ensemble with temperature control using a Nosé–Hoover thermostat \cite{nose2002molecular}. The time step is set to 2 fs for hydrogen-free systems and reduced to 1 fs for simulations including hydrogen.

\subsection{Deposition simulations}

\begin{figure*}[t!]
    \centering
    \includegraphics{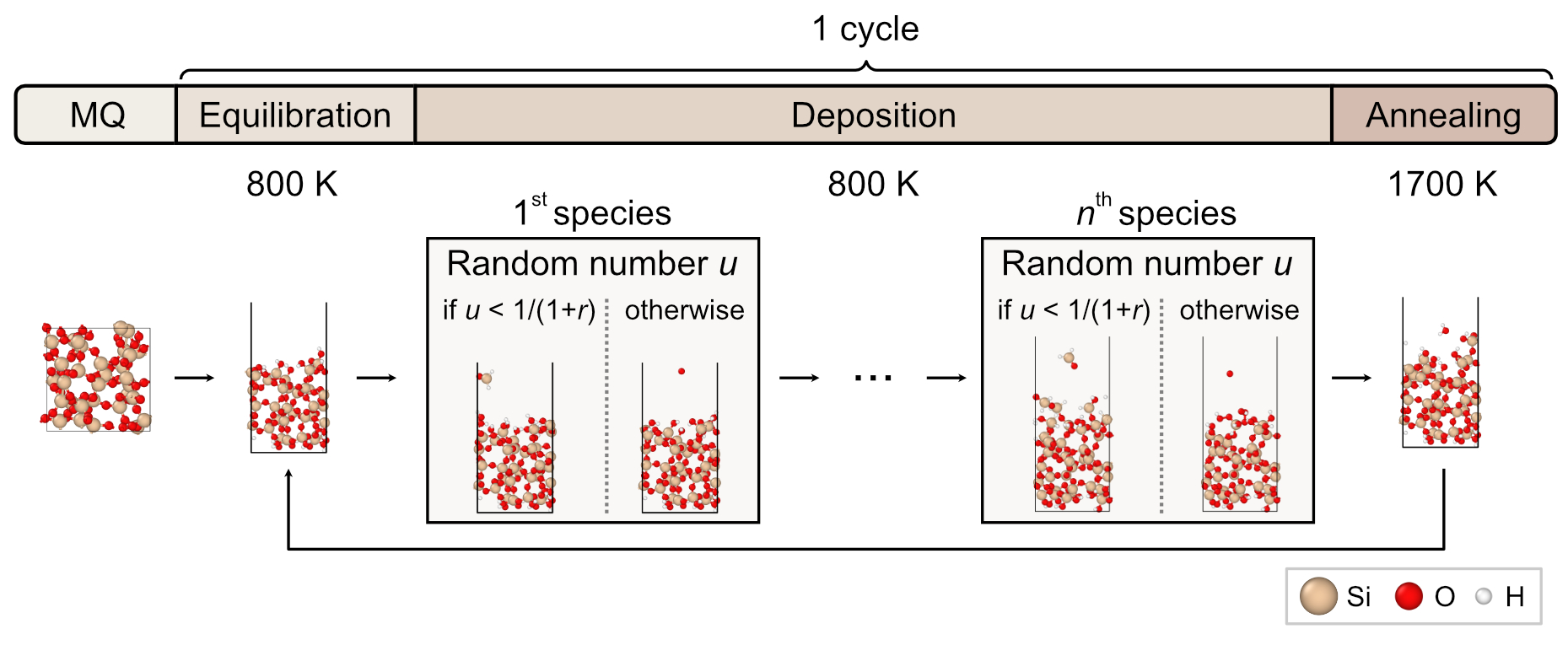}
    \caption{PECVD simulation protocol consisting of amorphous bulk generation, surface equilibration, deposition, and annealing steps. One cycle of the SiO$_2$ PECVD simulation includes equilibration and deposition at 800~K, followed by annealing at 1700~K. For each impact, a uniform random number $u \in [0,1)$ is generated to stochastically select the incident species, thereby statistically realizing the target ratio $r = \mathrm{O}/\mathrm{SiH}_2\mathrm{O}$ over multiple incidences. After the $n^\mathrm{th}$ deposition step (the determination of $n$ is described in the main text), the slab undergoes annealing, after which the next cycle begins.}
    \label{fig:deposition_sequence}
\end{figure*}

As the MLIP for PECVD simulations, we employ a pretrained universal 7net-0 model \cite{SevenNet0_2024}, which is trained on the MPtraj dataset \cite{deng2023chgnet}. 7net-0 is built on the NequIP architecture and employs an equivariant graph neural network framework \cite{batzner20223}, demonstrating excellent accuracy across a wide range of chemical systems while providing good scalability for large-scale MD simulations on multi-GPU platforms \cite{park2024scalable}. To enhance its accuracy for deposition simulations, we fine-tune the model, as described in the following section. MD simulations for SiO$_2$ PECVD are carried out using the LAMMPS package \cite{thompson2022lammps} under the NVT ensemble with a Nosé–Hoover thermostat \cite{nose2002molecular}. For enhanced computational performance and memory efficiency, the FlashTP CUDA kernel is integrated into the convolution layers of SevenNet~\cite{lee2025flashtp}, resulting in an approximately threefold throughput improvement. Van der Waals (vdW) interactions are considered by Grimme’s D3 method \cite{grimme2010consistent}.  

The simulation workflow adopted for SiO$_2$ PECVD is illustrated in Fig.~\ref{fig:deposition_sequence}. An amorphous SiO$_2$ model is first generated using melt--quench simulations; the corresponding melt--quench procedure is provided in the caption of Fig.~S1. A slab model representing the surface of a substrate is then constructed by introducing a vacuum region more than 30~\AA{} above the amorphous structure. Dangling bonds of surface-exposed Si and O atoms are passivated by attaching OH and H species, respectively. The atomic positions of the slab model are fully relaxed and subsequently used for PECVD MD simulations. Periodic boundary conditions are applied only in the lateral directions. During the MD simulations, atoms within the bottom 5.5~\AA{} of the slab are fixed to prevent rigid-body drift. A constant 0.5 fs timestep applies throughout the MD simulations.

The PECVD simulation proceeds through repeated cycles of equilibration, deposition, and annealing. During the equilibration stage, the substrate is equilibrated for 10~ps in the NVT ensemble at a temperature of 800~K. This substrate temperature is set to be slightly higher than the typical experimental deposition temperature ($\sim$600~K)~\cite{batey1986low} to accelerate reaction kinetics. During the deposition stage, a plasma species is introduced 6~\AA{} above the surface with a random in-plane position and molecular orientation. It is assigned an initial kinetic energy of 0.05~eV, corresponding to approximately $k_{\mathrm{B}}T$ at 600~K, and then propagates toward the surface. After incidence, the system evolves for 10~ps in the NVE ensemble to allow energy transfer from the incident species to the substrate, followed by an additional 10~ps of equilibration in the NVT ensemble at 800~K. These deposition steps are repeated until the criterion associated with the number of incident species, described in detail below, is satisfied. Note that the 20~ps time interval between successive molecular incidences (10~ps in the NVE ensemble followed by 10~ps in the NVT ensemble) is sufficient to capture the essential reaction dynamics of incident species, given that plasma-surface interactions typically occur on picosecond time scales \cite{graves2009molecular}. Nonetheless, the plasma-species flux employed in our simulations ($5\times{}10^{24}$~cm$^{-2}$~s$^{-1}$) is significantly higher than reported values \cite{kushner1993plasma}, implying that long-term processes such as film densification involving slow atomic diffusion and byproduct production may not be completely reflected within the current MD protocol. To partly mitigate this limitation, the film obtained after each deposition stage undergoes a high-temperature NVT anneal at 1700~K for 1~ns prior to the subsequent simulation cycle.

\begin{figure*}[t!]
    \centering
    \includegraphics{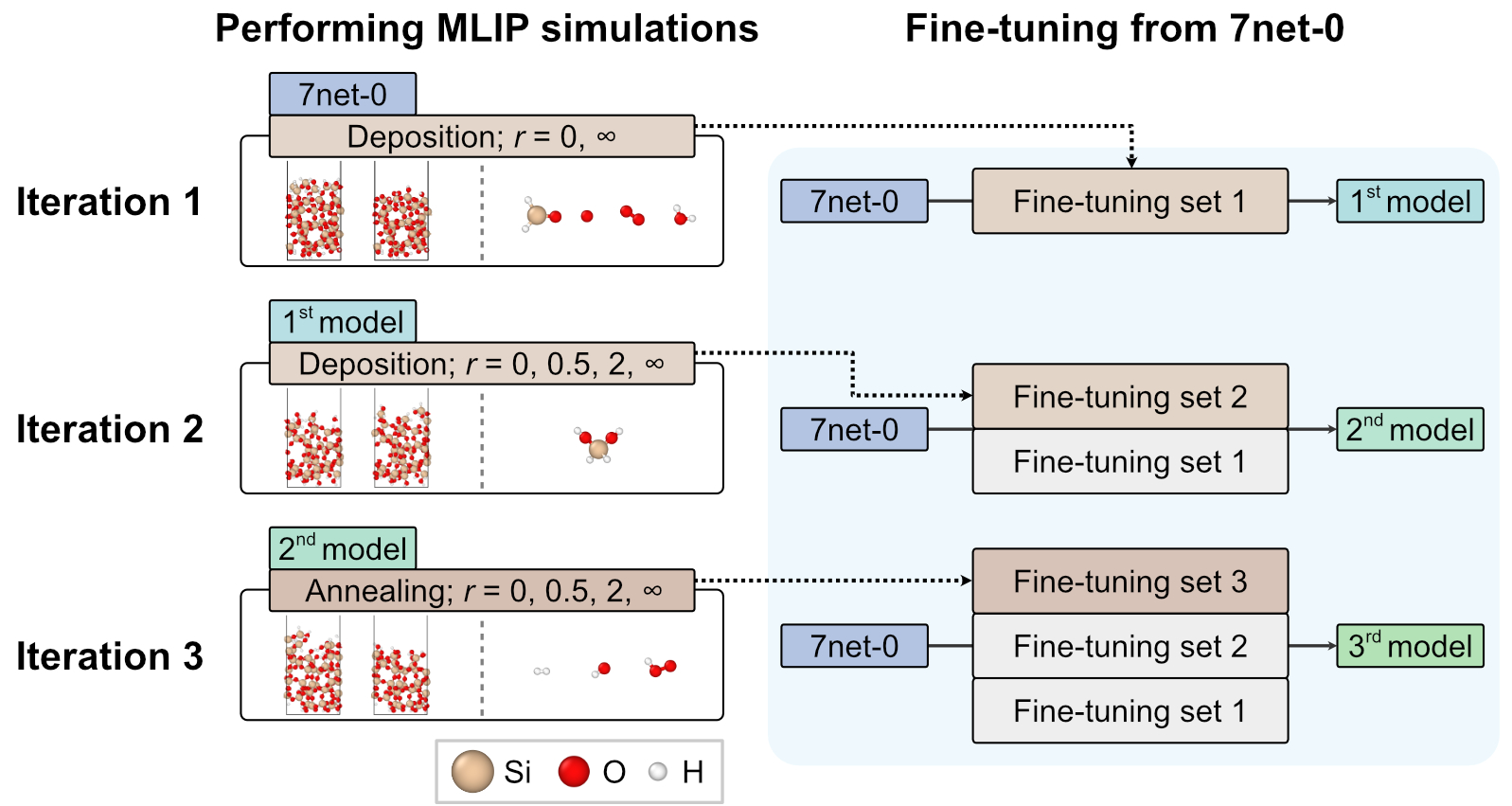}
    \caption{Procedure of iterative fine-tuning. In Iteration 1, deposition simulations at $r = 0$ and $\infty$ are performed using 7net-0 to generate Fine-tuning set 1. In Iteration 2, deposition simulations at $r = 0$, 0.5, 2, and $\infty$ are carried out using the model fine-tuned on Fine-tuning set 1 to produce Fine-tuning set 2. In Iteration 3, post-deposition annealing simulations at the same $r$ values are conducted using the model fine-tuned on the combined Fine-tuning sets 1 and 2, yielding Fine-tuning set 3. The final fine-tuned model is obtained by training on the cumulative dataset comprising Fine-tuning sets 1--3. The sampled configurations include both slab snapshots and gas-phase species extracted from the MD trajectories. Fine-tuning is performed by retraining 7net-0 on the cumulative fine-tuning dataset at each iteration.}
    \label{fig:schematic_mlip_training}
\end{figure*}

Among the various plasma species, we focus on SiH$_2$O as a silane-derived species and atomic oxygen as the oxidant, because they have been reported to play critical roles in SiO$_2$ PECVD processes \cite{kushner1993plasma,radouane2000rf,park2004low,zhang2024formation}. Although nitrogen-containing species (such as neutral NO and its electronically excited form) are also generated in the plasma phase, previous studies suggested that they have a minimal impact on film deposition under standard device fabrication conditions due to their limited residence times \cite{calta2024depth,del1993comparative,smith1993chemistry,wolf1995mass,martinu2010plasma}. Consequently, the nitrogen content in PECVD-grown SiO$_2$ films is usually negligible. The total number of incident plasma species in each SiO$_2$ deposition step is determined as $10(1+r)$~nm$^{-2}$ multiplied by the surface area of the supercell model (4 nm$^2$), where $r$ is the predefined ratio between the numbers of incident O atoms and SiH$_2$O molecules (i.e., $r = \mathrm{O}/\mathrm{SiH}_2\mathrm{O}$). The prefactor 10 represents the approximate number of SiH$_2$O molecules per unit surface area (nm$^{-2}$) required to convert all surface-exposed --OH species into Si--O bonds. The sequence of incident species (SiH$_2$O or O) is determined stochastically; a uniform random number $u \in [0,1)$ is first generated. If $u < 1/(1+r)$, a SiH$_2$O molecule is injected, whereas atomic oxygen is injected otherwise. This procedure produces a random arrival sequence with binomial fluctuations around the expected species counts based on $r$. In the present study, $r$ values of 0.25, 0.5, 1, 2, 3, 4, 5, and 6 are considered, and simulations are conducted for six PECVD simulation cycles.

\section{Results and discussion}

\subsection{Fine-tuning of 7net-0}

We first validate the performance of 7net-0. The model exhibits excellent accuracy for amorphous SiO$_2$ ($a$-SiO$_2$) bulk structures, yielding energetics and structural properties comparable to DFT results; as shown in Fig. S1, the radial distribution functions (RDFs) and angular distribution functions (ADFs) of $a$-SiO$_2$ models generated using DFT and 7net-0 show good agreement. These results demonstrate the strong transferability of the equivariant 7net-0 MLIP, given the fact that it has not been explicitly trained on amorphous systems \cite{kang2024graph}. 7net-0 also reasonably predicts the structures and energies of molecules associated with PECVD processes, including plasma species and possible byproducts such as H$_2$O, H$_2$, and O$_2$, as shown in Fig.~S2a. However, a non-negligible energy error is observed for atomic oxygen (0.26~eV/atom). In addition, 7net-0 predicts a molecular structure of SiH$_2$O that deviates from the DFT result as shown in Fig.~S2b. We also evaluate the accuracy for surface configurations that may occur during deposition. To this end, we perform PECVD simulations using 7net-0 at species ratios of $r = 0$ and $\infty$ at 1000~K. Some configurations exhibit non-negligible errors, particularly large force deviations, compared to the reference DFT calculations (Fig.~S3). Based on this comprehensive assessment, we conclude that further improvement of 7net-0 is required to achieve reliable PECVD simulations.

\begin{figure*}[t!]
    \centering
    \includegraphics{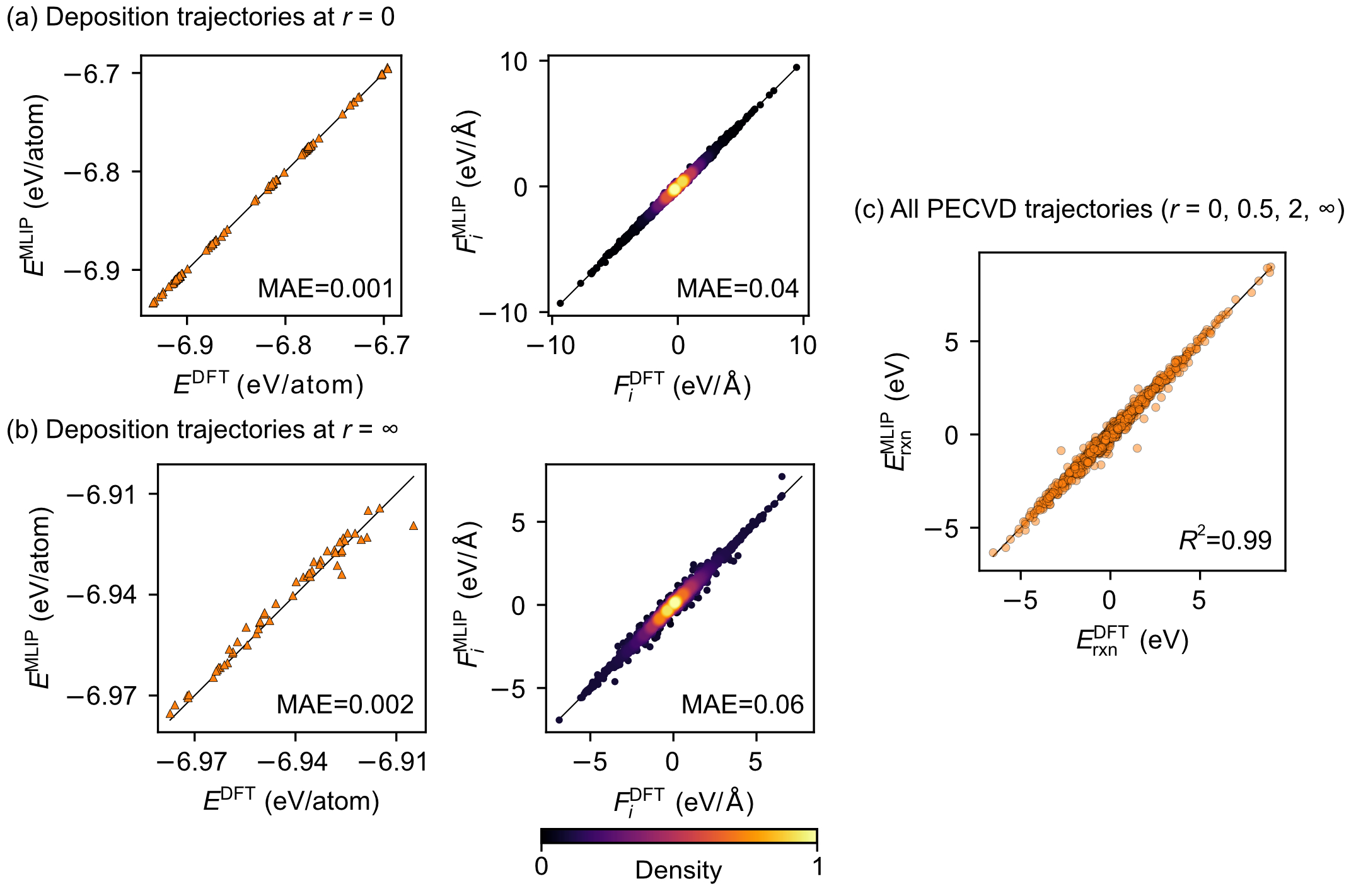}
    \caption{Validation of FT-MLIP. Parity plots comparing DFT and MLIP total energies (left) and Cartesian force components $F_i$ ($i \in \{x, y, z\}$) (right) for deposition trajectories at (a) $r = 0$ and (b) $\infty$. (c) Comparison of DFT and MLIP reaction energies ($E_\mathrm{rxn}$) extracted from MD trajectories, including deposition and annealing stages at $r = 0$, 0.5, 2, and $\infty$.}
    \label{fig:validation_simulation_FT}
\end{figure*}

To enhance the accuracy of 7net-0, we fine-tune the model on atomic configurations exhibiting large prediction errors greater than 5~meV/atom in energy or 0.5~eV/\AA{} in force. These thresholds are determined based on the energy and force errors observed in the validation of 7net-0 (orange dashed lines in Fig.~S3). We perform fine-tuning through multiple iterative stages (three iterations in this work) to construct a robust MLIP model \cite{ju2025application,an2025atomistic,kim2025neural,kim2025efficient,cho2026atomistic}. To produce the training dataset for fine-tuning at each iteration, we sample MD trajectories from deposition and annealing simulations performed using 7net-0 in the first stage and updated MLIP models in the second and third stages, as shown in Fig.~\ref{fig:schematic_mlip_training}. All MD simulations for dataset generation are carried out using small surface models containing approximately 100--200 atoms with a surface area of $1$~nm$^{2}$ to ensure the feasibility of DFT evaluations. To minimize redundancy in the training data, snapshots exhibiting weak structural correlations are selected based on graph-based analysis (Fig.~S4). In Iteration~1, we target configurations encountered during deposition at the extreme O/SiH$_2$O ratios $r=0$ and $\infty$ at 1000~K. Iteration~2 expands the sampling to deposition configurations over a broader range of species ratios, including $r=0$, 0.5, 2, and $\infty$, at the same temperature. Iteration~3 focuses on configurations produced during post-deposition annealing, in which deposited films at each $r$ are heated to 2000~K for 0.5~ns. The deposition (1000 K) and post-annealing (2000 K) temperatures used for generating the fine-tuning dataset are set higher than those employed in the actual PECVD simulations (800 K for deposition and 1700 K for post-annealing). The elevated temperature settings facilitate sampling of a broader and more diverse set of configurations \cite{an2025atomistic,kim2025neural}, which is likely to cover the configurational space relevant to PECVD simulations at lower temperatures and thereby enhance simulation stability. For the deposition simulations at each iteration, two independent trials are performed for each surface model, each involving six incident species, resulting in a total of 12 impact events for each value of $r$. Further increases in the number of trials or surface models are found to have a negligible impact on the fine-tuning results, as shown in Fig.~S5. 

In the fine-tuning dataset at each iteration, we also include MD trajectories of isolated plasma species and gaseous byproducts. This inclusion prevents undesirable shifts in their reference energies that may arise if only those embedded in the deposition supercell are considered (Fig.~S6). We perform 2-ps FPMD simulations at 1000~K, and then snapshots are sampled from the final 1~ps of each trajectory at intervals of 100~fs. Because isolated atomic oxygen has only one configuration owing to the absence of chemical bonds, we augment its representation in the fine-tuning dataset by duplicating this configuration to ensure accurate learning of its reference atomic energy.

All fine-tuned models at each iteration are built by updating the pretrained 7net-0 weights, rather than by consecutively updating weights across iterations. This fine-tuning approach starting from the 7net-0 weights can help preserve the pretrained knowledge and mitigate potential forgetting issues such as the loss of accuracy for bulk $a$-SiO$_2$ structures, which may occur during fine-tuning through multiple stages. The fine-tuning dataset is accumulated across iterations to enable incremental updates of chemical knowledge. The training for fine-tuning proceeds for 600 epochs using the Adam optimizer with a cosine-annealing learning-rate scheduler (50-step warm-up followed by a 200-step decay cycle). The learning rate reaches a maximum of $10^{-4}$ and decays to zero. We employ the Huber loss function with $\delta = 0.01$, applied to three targets: the total energy normalized per atom (eV/atom), all Cartesian force components for every atom (eV/\AA{}), and the six stress-tensor components (kbar). The force and stress terms are weighted by 0.1 and 0.001, respectively. After three iterations, the total number of accumulated configurations for fine-tuning is 1579, comprising 1459 surface structures and 120 molecular configurations. The dataset is randomly shuffled and split into training and validation subsets with a 9:1 ratio. On the validation set, the final fine-tuned MLIP after the third iteration (hereafter denoted as FT-MLIP) achieves significantly low mean absolute errors (MAEs) of 1.5~meV/atom, 0.05~eV/\AA{}, and 0.19~kbar for the energy, force, and stress components, respectively~\cite{cvitkovich2024machine,liu2025generalized,choi2025atomistic}.

\begin{figure*}[t!]
    \centering
    \includegraphics{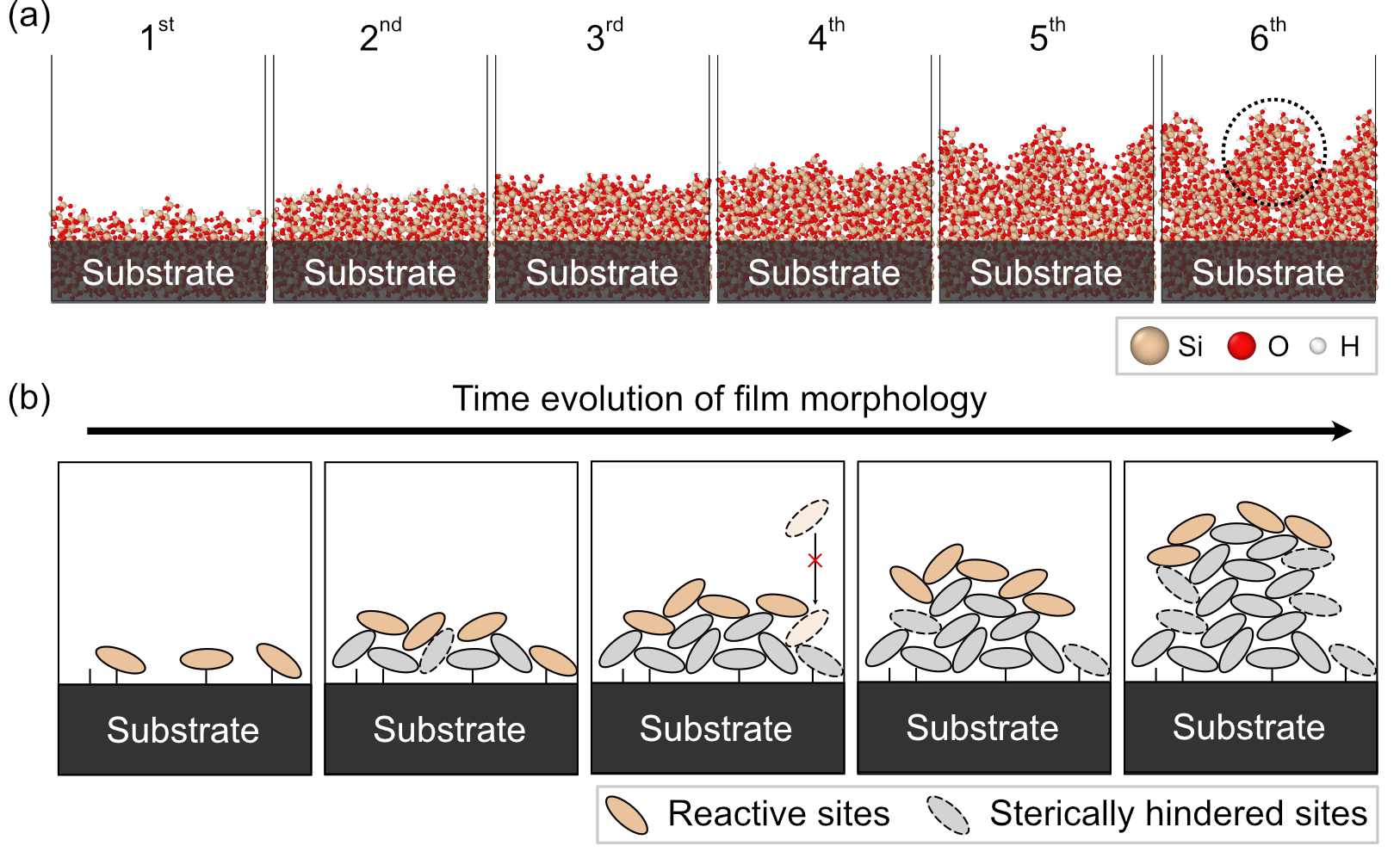}
    \caption{(a) Structural evolution during successive deposition cycles, showing progressive film thickening and the emergence of island-like growth (dotted circle). The supercell is expanded twice along the lateral direction for clarity. (b) Schematic illustration of how steric hindrance and prompt chemisorption of plasma species promote localized growth over time.}
    \label{fig:schematic_growth}
\end{figure*}

\begin{figure}[t!]
    \centering
    \includegraphics{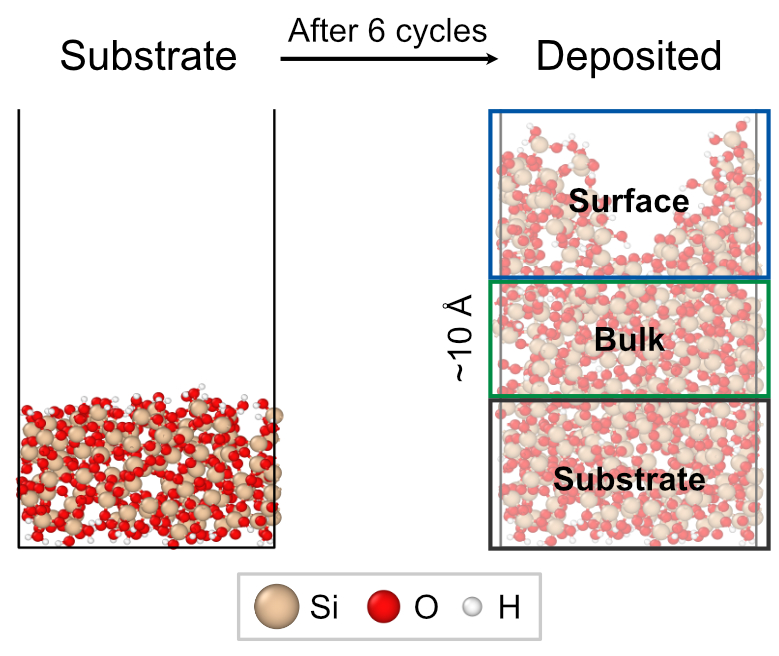}
    \caption{Atomic structures before deposition (left) and after six deposition cycles (right) at $r = 6$. The black, green, and blue lines denote the boundaries of the substrate, deposited bulk region, and surface region, respectively.}
    \label{fig:deposited_film_example}
\end{figure}

We confirm that FT-MLIP preserves the predictive accuracy for systems already well captured by the 7net-0 model, such as $a$-SiO$_2$ (Fig.~S1). It also exhibits improved performance for gaseous byproduct species (Fig.~S7). For validation purposes, PECVD simulations are performed using FT-MLIP on a small SiO$_2$ surface model (1~nm$^{2}$) at 1000~K for deposition and 2000~K for post-deposition annealing, with $r = 0$, 0.5, 2, and $\infty$. It turns out that FT-MLIP yields configurations with relatively small MAEs, specifically 0.001~eV/atom for energy and 0.04~eV/\AA{} for force at $r = 0$, and 0.002~eV/atom for energy and 0.06~eV/\AA{} for force at $r = \infty$ (Fig.~\ref{fig:validation_simulation_FT}a and b, respectively). It is worth noting that, compared to 7net-0, FT-MLIP exhibits improved accuracy and does not generate configurations with the pronounced energy and force errors observed during the 7net-0 validation. We also examine the energy and force parity for $r=0.5$ and $2$ (Fig.~S8), confirming the high accuracy of the FT-MLIP model for other $r$ values. Fig.~\ref{fig:validation_simulation_FT}c presents reaction energies associated with the reorganization of chemical bonds during the deposition and annealing stages for $r = 0$, 0.5, 2, and $\infty$. We employ a graph-based method to identify reaction events from the FT-MLIP MD trajectories, and the reaction energy is evaluated as the energy difference between the configurations before and after each chemical reaction. Excellent agreement with DFT is obtained, yielding $R^2 = 0.99$, and the MAE is improved by approximately 25$\%$ compared to the original 7net-0 model, further highlighting the suitability of the FT-MLIP for PECVD simulations. In the following sections, we present SiO$_2$ PECVD simulations performed using FT-MLIP.

\subsection{Growth behavior}

Fig.~\ref{fig:schematic_growth}a demonstrates the temporal evolution of the film morphology during PECVD simulations at $r = 6$. (Similar trends are also found across the examined $r$ values.) We observe that both SiH$_2$O and atomic oxygen readily adsorb onto the substrate through chemical reactions with surface groups such as Si--OH and Si--H species. Due to their high reactivity, the plasma species preferentially undergo immediate chemisorption upon impact, instead of transient physisorption. 

This rapid chemisorption behavior plays a critical role in determining the morphology of the deposited film. Specifically, as deposition proceeds, incident SiH$_2$O molecules react immediately at their impact sites due to their high reactivity, forming new surface species locally. These deposited species partially shield the underlying surface, reducing the accessibility of lower-lying adsorption sites to subsequent plasma species through steric hindrance (Fig.~\ref{fig:schematic_growth}b). As a consequence, growth becomes localized to specific regions, as highlighted by the circled area in Fig.~\ref{fig:schematic_growth}a. This growth mode suppresses uniform layer-by-layer deposition and instead promotes localized material accumulation. As a result, film growth becomes increasingly non-uniform, giving rise to pronounced surface roughness during PECVD. This mechanistic picture provides an atomic-scale explanation for the significant surface roughness commonly observed in PECVD-grown SiO$_2$ films \cite{park2004low}. 

\begin{figure}[t!]
    \centering
    \includegraphics{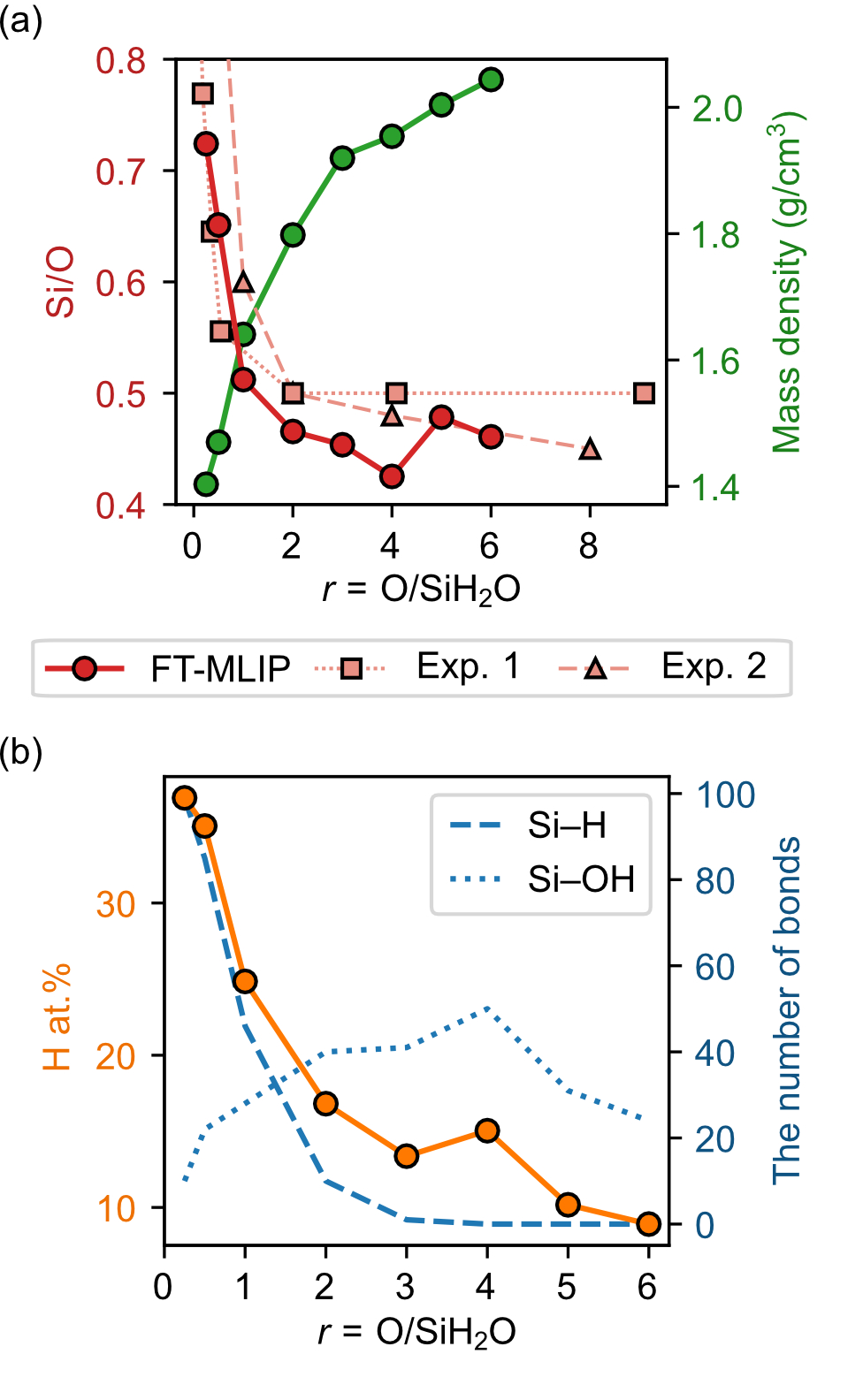}
    \caption{(a) Si/O ratio (left axis) and mass density (right axis) of the deposited bulk region (structures optimized at 0~K). Experimental data are mapped by aligning the N$_2$O/SiH$_4$ flow ratio in experiments with the simulated $r$ values, followed by appropriate scaling. Experimental data sets, Exp.~1 and Exp.~2, are taken from Refs.~\citenum{deenapanray1998characterization} and \citenum{calta2024depth}, respectively. (b) Hydrogen content (left axis) and the number of Si--H and Si--OH bonds in the bulk region (right axis) as functions of $r$.}
    \label{fig:properties}
\end{figure}

\subsection{Structural properties of the deposited films}

We present in Fig.~\ref{fig:deposited_film_example} the structures of the SiO$_2$ substrate and the deposited films formed on top of it for $r = 6$ after six deposition cycles (the corresponding structures for other $r$ values are shown in Fig.~S9). A bulk-like region with a thickness of approximately 10~\AA{} is clearly observed above the substrate. The local atomic structure of the bulk-like region of the deposited film is characterized by calculating the RDF and ADF for all $r$ values (Fig.~S10). Overall, the positions of the main RDF and ADF peaks for the deposited films reasonably coincide with those of amorphous bulk SiO$_2$, indicating the formation of SiO$_4$ tetrahedral amorphous networks.

We evaluate the Si/O ratio and mass density of deposited SiO$_2$ for the bulk region as a function of $r$, as shown in Fig.~\ref{fig:properties}a. The results show that the Si/O ratio notably decreases with increasing $r$ for $r<2$, after which the ratio is slowly varying. The decrease in the Si/O ratio originates from enhanced oxidation reactions owing to a higher probability of oxygen impact. Specifically, at low $r$ values, surface Si--H bonds are readily buried within the growing film due to the limited availability of oxidants, resulting in relatively high Si/O ratios. However, as $r$ increases, the higher oxygen flux more effectively oxidizes Si--H bonds to form Si--OH species (Fig.~\ref{fig:properties}b), which subsequently condense to form Si--O--Si linkages (the detailed reaction pathways are discussed in the following subsection). As a result, the Si/O ratio of the deposited film is gradually reduced. 

We compare this trend with experimental results \cite{deenapanray1998characterization,calta2024depth} by aligning the experimental gas flow ratio N$_2$O/SiH$_4$, at which the Si/O ratio begins to saturate, with the simulated plasma-species ratio $r = 2$, where a similar saturation trend appears, and then rescaling the N$_2$O/SiH$_4$ ratio accordingly. Note that a higher N$_2$O/SiH$_4$ flow ratio corresponds to a higher abundance of atomic oxygen in the plasma \cite{smith1993chemistry}. The scaled experimental curves (red dashed and dotted lines) exhibit composition trends consistent with our simulation results (red solid line), indicating that the simulated deposition process captures the essential experimental behavior. Previous FT-IR measurements also confirm that O-rich plasma phases facilitate the conversion of Si--H bonds into Si--OH species~\cite{zhang2024formation}.

On the other hand, the Si/O ratio obtained in our simulations after saturation ($r > 2$) slightly underestimates the ideal value of 0.5 for stoichiometric SiO$_2$. A similar trend is also often observed in experiments depending on the specific PECVD conditions (Fig.~\ref{fig:properties}a)~\cite{deenapanray1998characterization,calta2024depth}. This deviation is associated with insufficient hydrogen elimination during deposition, resulting in a relatively high hydrogen content in the deposited films. In particular, a hydrogen atomic fraction of 10~at.$\%$ or higher is observed for $r > 2$, as shown in Fig.~\ref{fig:properties}b. Such a high hydrogen concentration indicates a high abundance of Si--OH bonds in the films, together with a small number of Si--H bonds for $r = 2$ and 3. Consequently, the Si/O ratio of the films falls below 0.5. Moreover, incomplete hydrogen removal also contributes to the reduced mass density compared with the bulk value of 2.2~g/cm$^{3}$ (Fig.~\ref{fig:properties}a), because the formation of Si--O--Si linkages required for densification is suppressed. Consistent with our simulations, experimentally deposited films also exhibit Si/O ratios smaller than 0.5 when they contain significant hydrogen concentrations \cite{del1993comparative}.

\subsection{Reaction mechanisms}

\begin{figure*}[t!]
    \centering
    \includegraphics{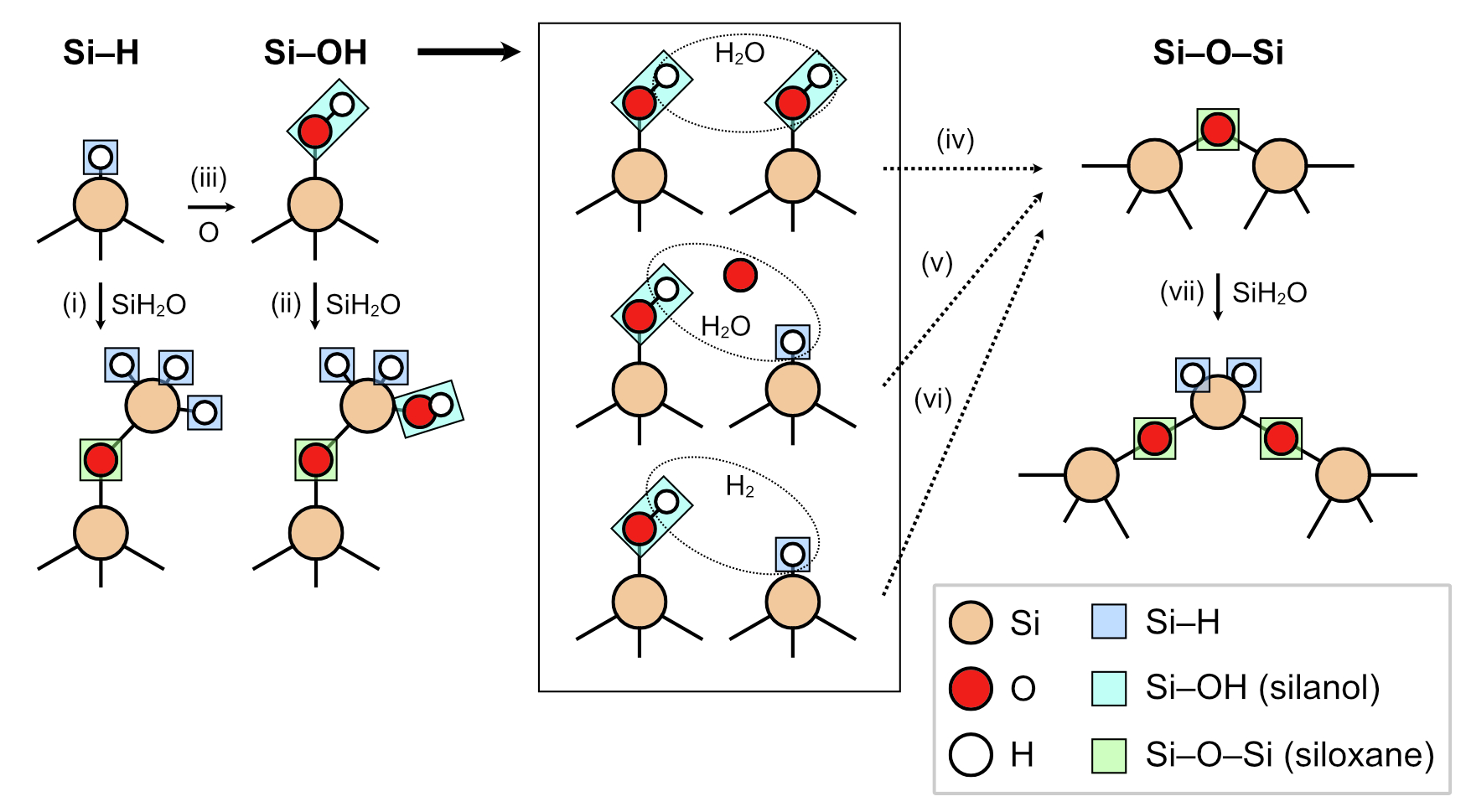}
    \caption{Schematic representation of major reaction pathways contributing to Si--O--Si network formation during PECVD. Reactions~\romannum{1}{} and \romannum{2}{} correspond to SiH$_2$O adsorption on surface Si--H and Si--OH terminations, respectively. Reaction~\romannum{3}{} denotes oxidation of Si--H. Reactions \romannum{4}{}-\romannum{6}{} describe the formation of Si--O--Si linkages through condensation processes involving hydrogen elimination. Reaction \romannum{7}{} represents SiH$_2$O adsorption on a surface Si--O--Si siloxane bridge.}
    \label{fig:reaction_mechanism}
\end{figure*}

\begin{figure}[t!]
    \centering
    \includegraphics{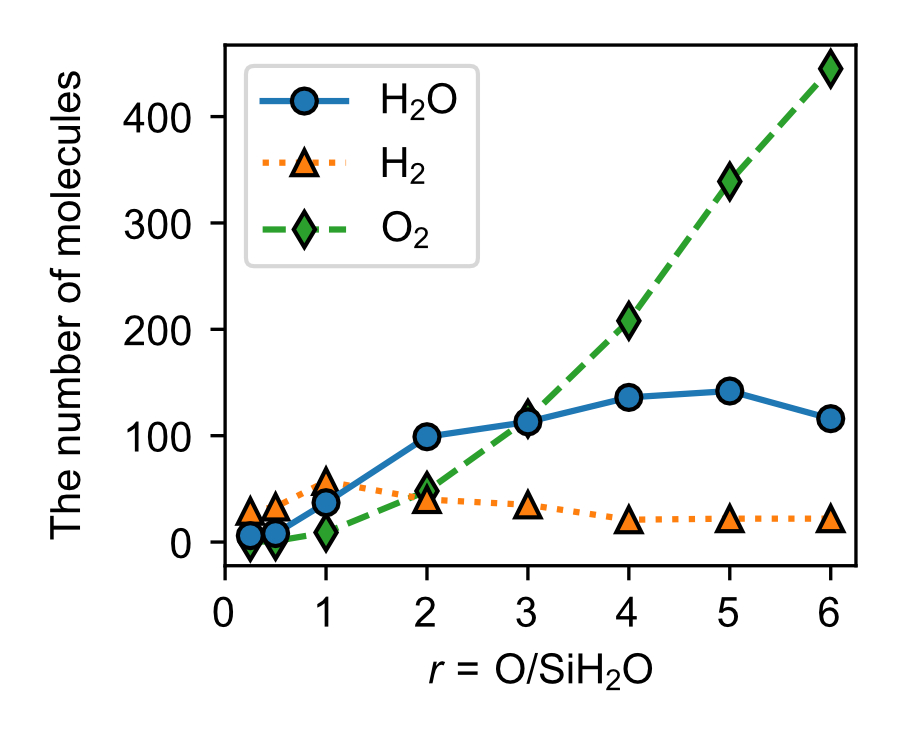}
    \caption{The number of gas-phase molecules generated during deposition and post-deposition annealing steps of PECVD simulations as a function of $r$.}
    \label{fig:major_byproducts}
\end{figure}

From the analysis of MD trajectories during PECVD simulations, we identify key reaction routes leading to the formation of the SiO$_2$ bonding network. We first discuss the major reaction pathways that are frequently observed regardless of $r$, as summarized in Fig.~\ref{fig:reaction_mechanism}. During PECVD, SiH$_2$O adsorbs on surfaces terminated with either Si--H or Si--OH groups. When SiH$_2$O adsorbs on a Si--H–terminated site, the hydrogen atom from the surface Si--H group is transferred to the adsorbate, while the oxygen atom in the adsorbate bridges the surface and incoming Si atoms (Reaction~\romannum{1}). On the other hand, when SiH$_2$O impinges on a Si--OH--terminated site, the hydrogen atom of the surface Si--OH is transferred to the oxygen atom of the adsorbate, leading to the formation of a Si--O--Si bridge (Reaction~\romannum{2}). These adsorption processes directly form Si--O--Si linkages, thereby contributing to SiO$_2$ growth. At the same time, SiH$_2$O adsorption generates new Si--H and Si--OH species on the surface. 

In parallel, incident atomic oxygen oxidizes surface Si--H bonds into Si--OH species (Reaction~\romannum{3}). As Si--OH groups accumulate on the surface, neighboring Si--OH pairs that are sufficiently close undergo condensation reactions, forming Si--O--Si linkages and releasing gaseous H$_2$O as a byproduct (Reaction~\romannum{4}). This condensation process plays a critical role in eliminating hydrogen from the growing film and in establishing a continuous, hydrogen-free Si--O--Si network within the SiO$_2$ film. In addition, atomic oxygen impinging between neighboring Si--OH and Si--H groups can also lead to the formation of a Si--O--Si linkage, producing an H$_2$O molecule (Reaction~\romannum{5}).

\begin{figure*}[t!]
    \centering
    \includegraphics{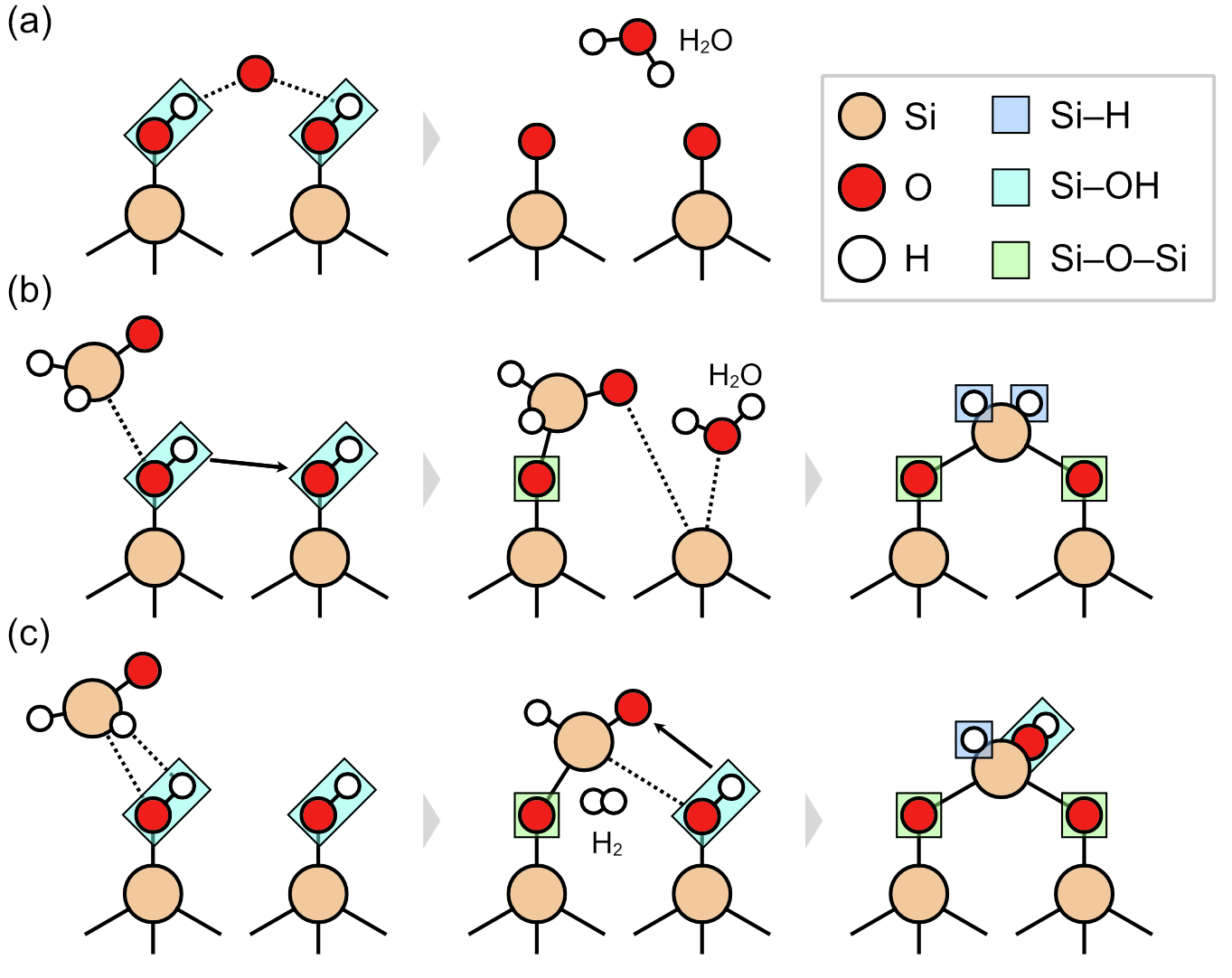}
    \caption{Schematic representation of secondary reaction pathways contributing to Si--O--Si network formation during PECVD. (a) H$_2$O formation due to oxidation of hydrogen on two neighboring Si--OH groups. (b,c) SiH$_2$O adsorption on two adjacent Si--OH groups, leading to the formation of (b) H$_2$O and (c) H$_2$ as byproducts.}
    \label{fig:reaction_mechanism_2}
\end{figure*}

Surface Si--OH and Si--H species can also directly interact when they are sufficiently close to each other (Reaction~\romannum{6}). This reaction forms a Si--O--Si bond and releases H$_2$, rather than H$_2$O, as a byproduct, significantly contributing to film growth at low $r$ values, as evidenced by the byproduct analysis in Fig.~\ref{fig:major_byproducts}. As summarized in Table S2, the PECVD process generates various gaseous species as byproducts, with $\text{H}_2\text{O}$, $\text{H}_2$, and $\text{O}_2$ identified as the major byproducts. For $r > 2$, these three species account for over 90\% of the total byproduct yield (Fig. S11). Notably, H$_2$ production is prominent at low $r$ values, exhibiting a level comparable to that of H$_2$O for $r < 2$. In contrast to H$_2$, the generation of H$_2$O, associated with Reactions~\romannum{4}{} and \romannum{5}{} discussed above, is consistently pronounced across the entire range of $r$ values examined in this study. 

Beyond this Si--H/Si--OH-mediated pathway, we also identify an additional reaction route involving Si--O--Si siloxane groups on the surface (Reaction~\romannum{7}). In this mechanism, incoming SiH$_2$O molecules interact directly with surface siloxane units, inserting into an existing Si--O bond and thereby extending the Si--O--Si linkage. The residual Si--H bonds generated during this insertion process subsequently undergo oxidation to form Si--OH species, which can further condense into Si--O--Si linkages via the pathway described above. Importantly, these reaction routes are observed across the full range of $r$ values examined in the present study, indicating that multiple surface-mediated mechanisms cooperatively govern SiO$_2$ network growth during PECVD.

Other secondary reaction pathways, although minor, also contribute to SiO$_2$ growth. First, incident atomic oxygen can react with hydrogen atoms residing on nearby Si--OH silanol groups for which direct self-condensation is kinetically unfavorable due to a large activation energy (Fig.~\ref{fig:reaction_mechanism_2}a). This reaction produces H$_2$O as a byproduct and leaves behind oxygen dangling bonds on the surface. These dangling bonds can facilitate hydrogen diffusion, which may increase the likelihood that initially distant silanol groups approach each other closely enough to undergo condensation into Si--O--Si linkages. The oxygen dangling bonds are eventually eliminated through reactions with atomic oxygen, contributing to the formation of Si--O--Si bonds and the release of O$_2$. 

\begin{figure*}[t!]
    \centering
    \includegraphics{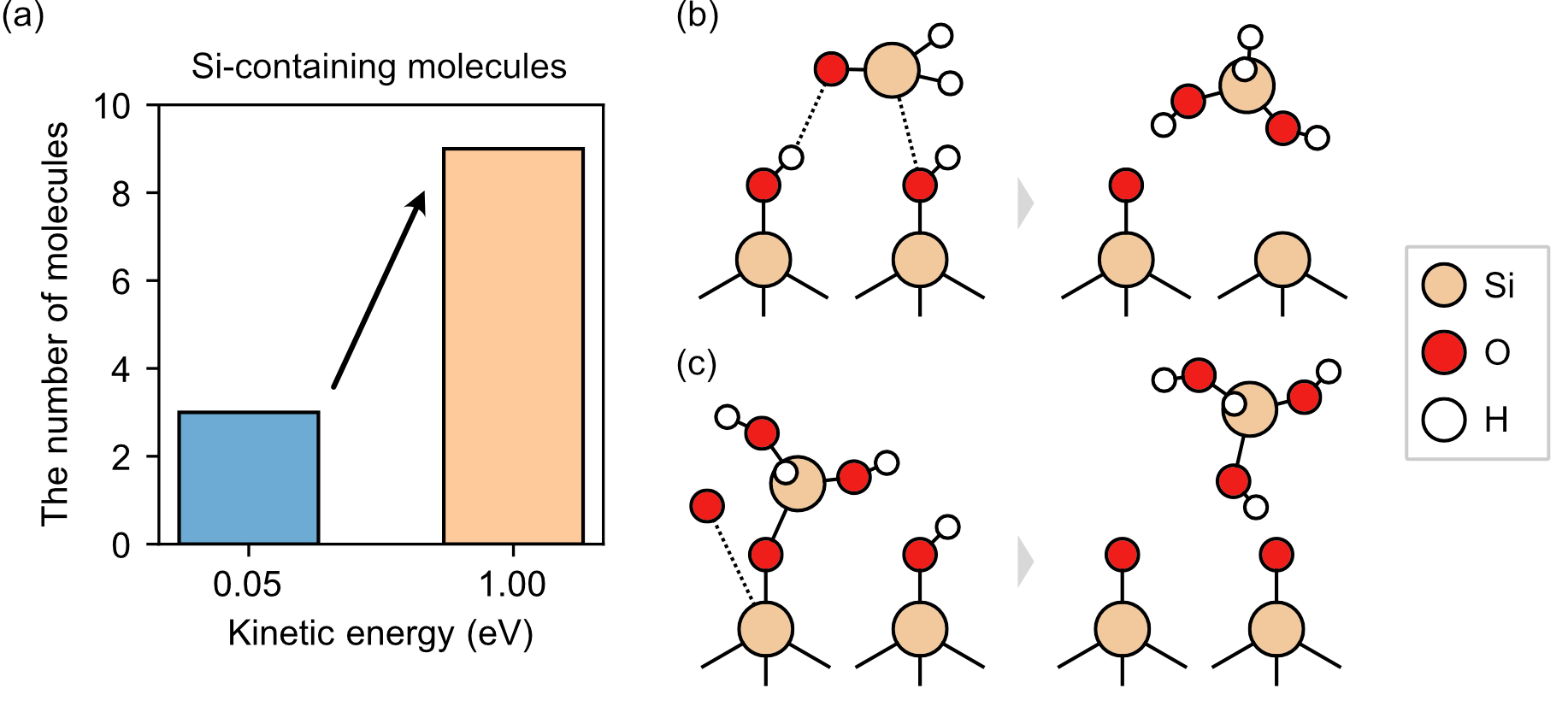}
    \caption{(a) The number of gas-phase molecules containing Si atoms generated during deposition steps of PECVD simulations as a function of kinetic energy of plasma species. (b,c) Schematic representation of reaction pathways generating Si-containing molecules during PECVD.}
    \label{fig:kinetic_energy_test}
\end{figure*}

SiH$_2$O adsorption on a silanol group can extend a Si--O--Si linkage when another silanol group is present nearby, as illustrated in Fig.~\ref{fig:reaction_mechanism_2}b. In this process, SiH$_2$O adsorption initially leads to hydrogen desorption from an oxygen atom of a silanol group, leading to the formation of a gaseous H$_2$O molecule. Subsequently, the oxygen atom in the adsorbate forms a chemical bond with a neighboring Si atom possessing a dangling bond. Similarly, when a hydrogen atom in SiH$_2$O approaches a hydrogen atom in a silanol group, H$_2$ formation can occur first, followed by the formation of a Si--O--Si linkage (Fig.~\ref{fig:reaction_mechanism_2}c). If an additional silanol group is present in the vicinity, the adsorbate can further promote the formation of another Si--O--Si bond, with the oxygen atom in the adsorbate accepting a hydrogen atom from the neighboring silanol group.

We also observe reactions that generate O$_2$ gas, which contribute little to SiO$_2$ growth. As $r$ increases, atomic oxygen impinges on the surface more frequently. As a result, silanol groups can undergo aggressive oxidation, rapidly forming hydroperoxy species such as Si--OOH and, in extreme cases, Si--OOOH (Fig.~S12). Similarly, siloxane Si--O--Si bridges can be converted into peroxo species, including Si--O--O--Si and Si--O--O--O--Si. The influence of these O-rich motifs on overall SiO$_2$ growth is limited; during subsequent oxygen impacts or thermal annealing, they readily revert to the corresponding Si--OH or Si--O--Si configurations, accompanied by the release of a gaseous O$_2$ molecule. Accordingly, at large $r$ values, O$_2$ becomes the predominant byproduct, as shown in Fig.~\ref{fig:major_byproducts}. 

\subsection{Impacts of experimental conditions}

Our simulation results and mechanistic analysis provide insights into the appropriate selection of temperature and RF power, two critical experimental parameters governing film growth, for optimizing SiO$_2$ PECVD using SiH$_4$ and N$_2$O precursor gases. Plasma activation facilitates chemical reactions necessary for SiO$_2$ formation. However, simply increasing the RF power to accelerate deposition by facilitating higher-energy plasma species and radicals (e.g., atomic oxygen) may degrade film quality, particularly by promoting excessive hydrogen incorporation, stoichiometric imbalance, and incomplete densification. Specifically, as demonstrated above, the oxidation process based on exposure to atomic oxygen alone is insufficient to fully eliminate hydrogen from the growing film. Instead, effective hydrogen removal requires condensation reactions between surface species, which primarily produce H$_2$O and, to a lesser extent, H$_2$. The efficiency of these condensation reactions is strongly temperature dependent. Therefore, under high RF-power conditions, although the conversion of Si--H into Si--OH species becomes more efficient, the resulting Si--OH groups may be quickly buried beneath subsequently deposited layers before undergoing condensation if the temperature is not high enough to activate corresponding reactions. In this regard, increasing the deposition temperature, within the limits imposed by device compatibility, together with careful control of RF power, is critical for improving film quality, including enhanced densification and reduced hydrogen incorporation. At the same time, the RF power must be optimized to balance the deposition rate and film quality. Indeed, Experiment~1 in Fig.~\ref{fig:properties}a, conducted at a higher temperature (573~K) and lower RF power (20~W) than Experiment~2 (523~K and 50~W), yields films of higher quality with improved stoichiometric balance. Moreover, previous literature corresponding to Experiment~1~\cite{deenapanray1998characterization} also reported lower-density films when a lower deposition temperature of 473~K was applied, further highlighting the importance of properly controlling temperature conditions.

In addition, although the present study focuses on neutral plasma species, a substantial population of ions can be generated in the plasma phase and reach the substrate with kinetic energies exceeding a few electronvolts under high RF-power conditions. These energetic ionic species can interact strongly with the growing surface and may induce unintended etching effects. To examine this possibility, we increase the kinetic energy of the incident species from 0.05 to 1.0 eV at $r = 4$ to reflect higher plasma-power conditions. The simulations reveal that a larger number of Si-containing gaseous species, particularly SiH$_n$(OH)$_{4-n}$, are produced during deposition compared with the baseline case (Fig.~\ref{fig:kinetic_energy_test}a). We identify several pathways for this etching process. First, high-energy SiH$_2$O molecules can directly generate volatile Si-containing species, mainly SiH$_n$(OH)$_{4-n}$, upon impact (Fig.~\ref{fig:kinetic_energy_test}b and Fig.~S13a, c). Second, the incidence of high-energy atomic oxygen can produce Si-containing volatile species by modifying the chemical bonding of Si on the surface (Fig.~\ref{fig:kinetic_energy_test}c and Fig.~S13b). While some of these volatile Si-containing species desorb immediately, others may remain near the surface for a relatively long time due to strong physisorption mediated by hydrogen bonding. However, the latter also eventually desorbs through energy transfer from subsequent impacts. These etching processes would limit the growth rate and contribute to the increased surface roughness under elevated RF-power conditions, which was indeed reported in experiments~\cite{park2004low}.

\section{Conclusion}

In this work, we investigated the atomic-scale mechanisms of SiO$_2$ PECVD using MD simulations driven by an MLIP. By systematically varying the oxidant-to-SiH$_2$O ratio $r$, we clarified how the precursor supply influences film composition, density, and hydrogen incorporation. The simulations reveal that Si--O--Si network formation primarily proceeds through oxidation of surface Si--H species to form Si--OH groups, followed by condensation reactions between neighboring Si--OH groups that produce H$_2$O as the dominant byproduct. At low $r$ values, H$_2$ formation via reactions between Si--H and Si--OH species also contributes to network formation to some extent. Increasing oxidant supply promotes formation of the Si--O--Si network by oxidizing residual Si--H species, thereby reducing hydrogen incorporation and leading to saturation of the Si/O ratio. The simulations further show that rapid chemisorption of reactive plasma species and steric hindrance from pre-deposited species result in localized growth, which leads to non-uniform film morphology and surface roughness during PECVD. In addition, high-energy impacts of plasma species can generate volatile Si-containing fragments, indicating surface etching that may limit growth rates and enhance surface roughness under high RF-power conditions. Overall, this study provides atomistic insights into PECVD growth mechanisms which are important for optimizing SiO$_2$ PECVD processes. Moreover, the computational approach and analytical methodologies established herein are broadly applicable to other thin-film deposition processes, thereby contributing to the advancement of future thin-film-based technologies.

\section*{CRediT authorship contribution statement}
\textbf{J. Kim:} Writing -- original draft, Conceptualization, Methodology, Investigation, Validation, Formal analysis, Data curation. \textbf{M. Moon:} Investigation, Methodology. \textbf{H. Cho:} Investigation, Methodology. \textbf{H.-D. Kim:} Project administration, Funding acquisition. \textbf{R. Kim:} Funding acquisition. \textbf{G. Park:} Funding acquisition. \textbf{S. Han:} Supervision. \textbf{Y. Kang:} Writing -- review \& editing, Conceptualization, Supervision, Methodology.

\section*{Declaration of competing interest}
The authors declare that they have no known competing financial interests or personal relationships that could have appeared to influence the work reported in this paper.

\section*{Acknowledgments}

This work was supported by Samsung Display Company Ltd and the National Research Foundation of Korea (NRF) funded by the Korea government (MSIT) (RS-2025-25442273).

\section*{Data Availability}
Training data for the MLIPs will be available upon request, except for the open database. The SevenNet code used in this study is available in the GitHub repository: \href{https://github.com/MDIL-SNU/SevenNet}{https://github.com/MDIL-SNU/SevenNet}.

\bibliographystyle{elsarticle-num} 
\biboptions{numbers,sort&compress}
\bibliography{reference.bib}

\end{document}